\documentclass[pre,floatfix,twocolumn,showpacs]{revtex4}
\usepackage{amsmath}
\usepackage{amsfonts}
\usepackage{epsfig}

\begin{document}
\title{Modular networks emerge from multiconstraint optimization}

\author{Raj Kumar Pan}
\author{Sitabhra Sinha}
\affiliation{%
The Institute of Mathematical Sciences, C.I.T. Campus, Taramani,
Chennai - 600 113 India
}%
\date{\today}

\begin{abstract}
Modular structure is ubiquitous among complex networks. We note that most
such systems are subject to multiple structural and functional constraints,
e.g., minimizing the average path length and the total number of links,
while maximizing robustness against perturbations in node activity. We show
that the optimal networks satisfying these three constraints are
characterized by the existence of multiple subnetworks (modules) sparsely
connected to each other. In addition, these modules have distinct hubs
resulting in an overall heterogeneous degree distribution.
\end{abstract}
\maketitle

Complex networks have recently become a focus of scientific attention, with
many natural, social and technological networks seen to share certain
universal structural features~\cite{Newman03,Albert02}. These networks
often exhibit topological characteristics that are far from random.  For
instance, they show a significant presence of {\em hubs}, i.e., nodes with
large degree or number of connections to other nodes. 
Indeed, hubs are crucial for
linking the nodes in real networks, which have extremely sparse connectivity,
with the probability of connection between any pair of nodes, $C$, varying
between $10^{-1}$ and $-10^{-8}$~\cite{Newman03}. By contrast, random networks 
with such small $C$ are almost always disconnected.  
The hubs also lead to the ``small-world'' effect~\cite{Watts98} by
reducing the average path length of the network. Another property observed
in many networks is the existence of a modular structure. We define a
network to be {\em modular} if it exhibits significantly more 
intramodular connections compared to intermodular connections. 
Such networks can be decomposed into distinct
subnetworks or modules by removing a few links.  
Modular networks observed in empirical studies span a wide
range from cellular networks involved in metabolism and
signalling~\cite{Holme03}, to cortical
networks~\cite{Hilgetag00}, social networks~\cite{Arenas04}, food
webs~\cite{Krause03} and the internet~\cite{Eriksen03}.  Many of these
networks also exhibit large number of hubs, which often have the role of
interconnecting different modules~\cite{Vespignani03}.

The majority of previous studies on modular networks have been concerned
with methods to identify community structure~\cite{Girvan02}.
There have been relatively few attempts to explain the potentially more
interesting question of how and why modularity emerges in complex networks.
Most such attempts are based on the notion of evolutionary pressure, where
a system is driven by the need for adapting
to a changing environment~\cite{Variano04,Kashtan05}. However, such
explanations involve complicated adaptive mechanisms, in which the
environment itself is assumed to change in a modular fashion. Further,
adaptation might lead to decrease in connectivity through biased selection
of sparse networks, which eventually results in {\em disruption of the
network} with the modules being isolated nodes~\cite{Variano04} or
disconnected parts~\cite{Sole03}. More recently, a social network
model has shown the emergence of isolated communities
through the rearrangement of links to form groups with
homogeneous opinion~\cite{Holme06}. 

A crucial limitation of these above studies is that they almost always
focus on a single performance parameter. However, in reality, most
networks have to optimize between several, often conflicting, constraints.
While structural constraints, such as path length, had been the focus of
initial work by network researchers, there has been a growing realization
that most networks have dynamics associated with their
nodes~\cite{Strogatz01}.  The robustness of network behavior
is often vital to the efficient functioning
of many systems, and also imposes an important constraint on networks.
Therefore, the role played by dynamical considerations in determining the
topological properties of a network is a challenging and important question
that opens up new possibilities for explaining observed features of complex
networks~\cite{Guimera02}. In this letter, we propose a simple mechanism for the emergence
of modularity in networks as an optimal solution for satisfying a minimal
set of structural and functional constraints.  These essentially involve
(i)~reducing the average path length, $\ell$, of a network by (ii)~using a
minimum number of total links, $L$, while (iii)~decreasing the
instability of dynamical states associated with the network.

We investigate the dynamical stability of a network composed of $N$ nodes,
which are self regulating when isolated, by measuring the growth rate of a
{\em small} perturbation ${\bf x}$ about an equilibrium state of the
network dynamics. Although the system can be nonlinear in
general, the dynamics of such perturbations are described by a {\em linear}
system of coupled differential equations $\dot{x_i} = \sum_{j=1}^N J_{ij}
x_j$.  The stability of the equilibrium is then determined by the largest
real part $\lambda_{\text{max}}$ of the eigenvalues for the matrix $\mathbf{J}$
representing the interactions among the nodes. The perturbation decays if 
$\lambda_{\text{max}}<0$, and increases otherwise, at a rate proportional to
$|\lambda_{\text{max}}|$. Thus, minimizing $\lambda_{max}$ makes the 
equilibrium less unstable, which is important for many systems including
ecological networks~\cite{May73}. Here $J_{ii} = -1 ~\forall i$
such that we only consider instability induced through network
interactions.  The off-diagonal matrix elements $J_{ij} (\sim A_{ij}
W_{ij})$ include information about both the topological structure of the
network, given by the adjacency matrix $\mathbf{A}$ ($A_{ij}$ is $1$, if
nodes $i,j$ are connected, and $0$, otherwise; $A_{ii}=0 ~\forall i$), as
well as, the distribution of interaction strengths $W_{ij}$ between nodes.
In our simulations, $W_{ij}$ has a Gaussian distribution with zero mean and
variance $\sigma^2$; however, a nonzero mean does not qualitatively change
our results. For an Erd\"{o}s-Renyi (ER) random network, $\mathbf{J}$ is a
sparse random matrix, with $\lambda_\text{max}\sim \sqrt{N C \sigma^2}-1$,
according to the May-Wigner theorem~\cite{May73}. Therefore, increasing the 
system size $N$, connectivity $C$ or interaction strength $\sigma$, results in 
instability of the network. This result has been shown to be remarkably
robust with respect to various generalizations~\cite{generalizations}.
Further, for uniform coupling strength, $\lambda_{max}$ is inversely related
to the epidemic propagation threshold for the network~\cite{Wang03}, and
hence, minimizing $\lambda_{max}$ also makes the network more robust
against spreading of infection.

Networks are also subject to certain structural constraints. One of them is
the need to save resources, manifested in minimizing {\em link cost}, i.e.,
the cost involved in building and maintaining each link in a
network~\cite{Mathias01}.  This results in the network having a
small total number of links, $L$. However, such a procedure runs
counter to another important consideration of reducing
the average path length $\ell$, which improves the network {\em efficiency} by
increasing communication speed among the nodes~\cite{Latora01}.  The
conflict between these two criteria can be illustrated through the example
of airline transportation networks.  Although, fastest communication (i.e.,
small $\ell$) will be achieved if every airport is connected to every other
through direct flights, such a system is prohibitively expensive as every
route involves some cost in maintaining it.  In reality, therefore, one
observes the existence of airline hubs, which act as transit points for
passengers arriving from and going to other airports.

\begin{figure}
\begin{center}
  \includegraphics[width=0.9\linewidth]{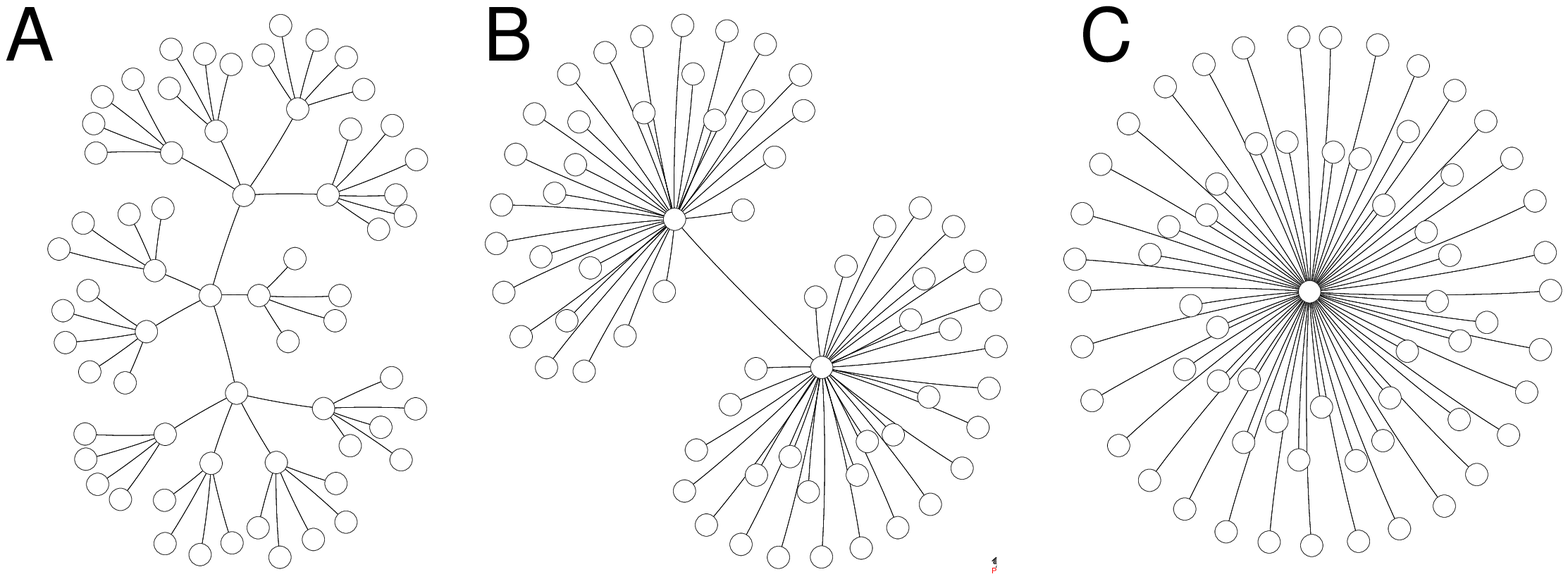}
  \includegraphics[width=0.9\linewidth]{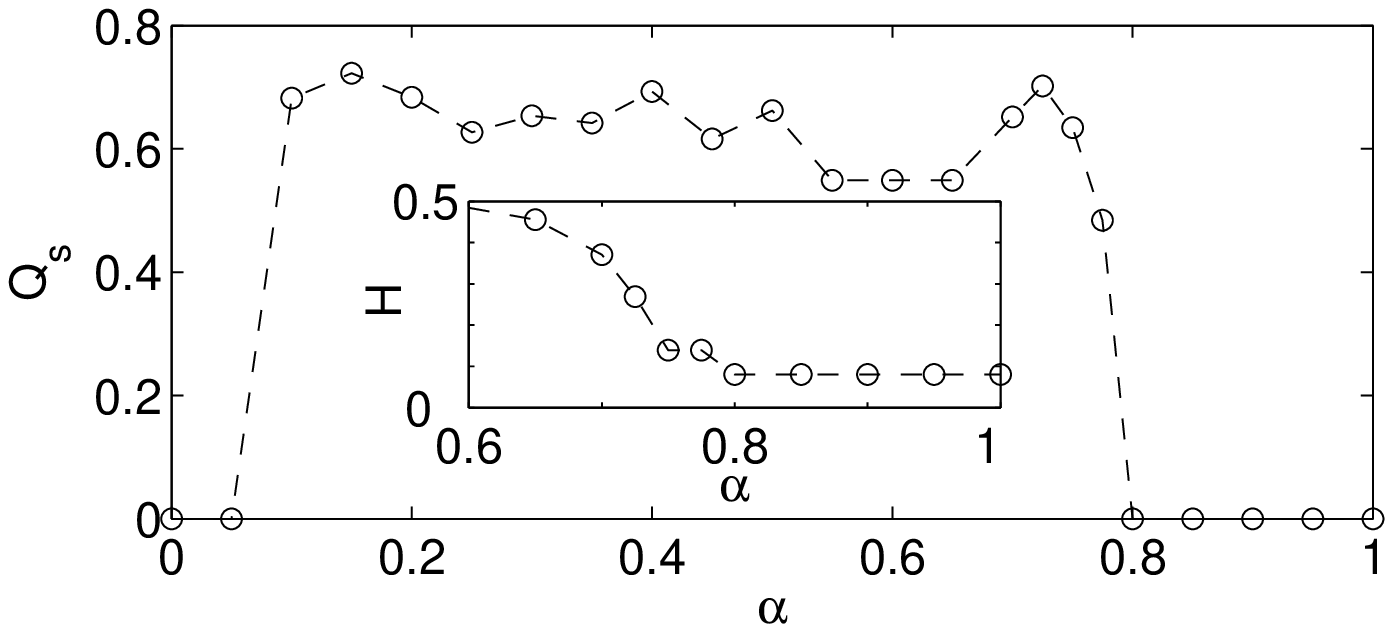}
  \end{center}
  \caption{The optimized network structures for a system with $N = 64$ 
  nodes and $L = N-1$, at different values of $\alpha$: (A) 0.4,
  (B) 0.775 and (C) 1. For $\alpha=0$ the optimal network is a 1-dim chain. 
  (Bottom) The modularity $Q_s$ of the optimized network for
  different $\alpha$, when each module is a community defined in the strong 
  sense. The transition to star configuration occurs around $\alpha \simeq 
  0.8$, as observed in the variation of degree entropy $H$ with $\alpha$.}
\label{fig_optimized}
\end{figure}

For ER random networks, although $\ell$ is low, $L$ is high because
of the requirement to ensure that the network is connected: $L > N
{\rm ln} N$~\cite{Bollobas01}.  Introducing the constraint of link cost
(i.e., minimizing $L$) while requiring low average path length
$\ell$, leads to a starlike connection topology
(Fig.~\ref{fig_optimized}C). A {\em star network} has a
single hub to which all other nodes are connected, there being no other
links.  Its average degree $\langle k \rangle \approx 2$ is {\em
non extensive} with system size, and is much smaller than 
a connected random network, where $\langle k \rangle \sim {\rm ln}~N$. 
However, such starlike networks are extremely unstable with respect to
dynamical perturbations in the activity of their nodes.  The probability of
dynamical instability in random networks increases only with average degree
($\lambda_{\text{max}} \sim \sqrt{\langle k \rangle}$, since $\langle k
\rangle=NC$), while for star networks it increases with the largest degree,
and hence the size of the network itself ($\lambda_{\text{max}} \sim
\sqrt{N}$). To extend this for the case of weighted networks we look at the
largest eigenvalue of $\mathbf{J}$, $\lambda_{max} =
-1+\sqrt{\sum_{i=2}^{N}J_{1i}J_{i1}}$,  the hub being labeled as node 1.
The stability of the weighted star network is governed by
$\sum_{i=2}^{N}J_{1i}J_{i1}$, which is the displacement due to a
1-dimensional random walk of $N-1$ steps whose lengths are products of
pairs of random numbers chosen from a $Normal~(0, \sigma^2)$ distribution.

To obtain networks which satisfy the dynamical as well as the structural
constraints we perform optimization using simulated annealing, with a
network having $N$ nodes and $N-1$ unweighted links (the smallest number
that keeps the network connected). Having fixed $L$, the energy
function to be minimized is defined as \[ E(\alpha) = \alpha \ell +
(1-\alpha)\lambda_{\text{max}},\] where the parameter $\alpha\in[0,1]$ 
denotes the relative importance of the path length constraint over
the condition for reducing dynamical instability. Rewiring is attempted 
at each step and is (i) rejected if the updated network is disconnected, (ii)
accepted if $\delta E=E_{final}-E_{initial}<0$, and (iii) if $\delta E >
0$, then accepted with probability $p=\exp(-\delta E/T)$, where $T$ is the
``temperature''. The initial temperature was chosen in such a way that
energetically unfavorable moves had $80 \%$ chance of being accepted.
After each monte carlo step ($N$ updates) the temperature was reduced by $1
\%$ and iterated till there was no change in the energy for 20 successive
monte carlo steps.   For each value of $\alpha$, the optimized network with
lowest $E$ was obtained from 100 realizations. 

As can be seen from Fig.~\ref{fig_optimized}, modularity emerges when the
system tries to satisfy the twin constraints of minimizing $\ell$ as well
as $\lambda_{\text{max}}$. When $\alpha$ is very high ($ \sim 0.8$) such
that the instability criterion becomes less important, the system
shows a transition to a starlike configuration with a single hub. However,
as $\alpha$ is decreased, the instability of the hub makes the star network less
preferable and for intermediate values of $\alpha$, the optimal network
gets divided into modules, as seen from the measure of network modularity,
$Q$~\cite{Newman04}. This is defined as
$Q=\sum_{s}[(L_s/L)-(d_s/2L)^2]$, where $L_s$ is the number of links 
between nodes within a module $s$, and $d_{s}$ is the sum of the
degrees of the nodes in $s$.
To obtain a robust partitioning of the network, we consider modules to be 
communities defined in the {\em strong} sense, i.e., each node $i$ belonging to a community
has more connections with nodes within the community than with the rest of
the network~\cite{Radicchi04}. The resulting
modularity measure $Q_s$ is high for a modular network, 
whereas for homogeneous, as well as, for starlike networks, $Q_s=0$. 
To determine the communities, we (1) compute the betweenness measure for all
edges and remove the one with highest score: (2a) if it results in 
splitting the network (or subnetwork) into communities in the strong sense, 
then the resulting $Q_s$ is computed; (2b) if not, we go back to step (1)
and remove the edge with the next highest score. 
The process is carried out iteratively until all edges of the network 
have been considered. Note that, in step (2a), checking whether
the splitting results in communities in the strong sense is considered with
respect to the full network. 
We verified these results by also calculating $Q_s$ 
with the network modules determined through stochastic extremal 
optimization~\cite{Duch05}.
The transition between modular and star
structures is further emphasised in the behavior of the degree entropy, $H
= - \sum_k p_k {\rm ln} p_k$, where $p_k$ is the probability of a node
having degree $k$. The emergence of a dominant hub at a critical value of
$\alpha$ is marked by $H$ reducing to a low value.

To understand why modular networks emerge on simultaneous
optimization of structural and functional constraints we look at the
change in stability that occurs when a star network is split into $m$
modules, the modules being connected through links between their
hubs. The largest eigenvalue for the entire system of $N$
nodes is the same as that for each isolated module, $\lambda_{\text{max}} \sim
\sqrt{N/m}$, as the additional effect of the few intermodular links is
negligible.  At the same time, the increase in the average path
length $\ell$ with $m$ is almost insignificant.  Therefore, by dividing the
network into a connected set of small modules, each of which is a star
subnetwork, the instability of the entire network decreases significantly
while still satisfying the structural constraints.

\begin{figure}
\begin{center}
  \includegraphics[width=0.9\linewidth]{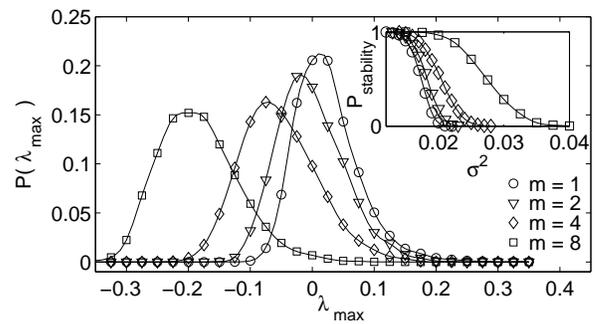}
\end{center}
\caption{Probability distribution of $\lambda_{\text{max}}$ for a clustered
star network ($N = 256, L = 15 N$) with different numbers of
modules, $m$. Modules of equal size are connected by single link
between respective hubs. Link weights $W_{ij}$ follow a
$Normal~(0,\sigma^2)$ distribution with $\sigma^2=0.018$. (Inset)
Probability of stability [$P(\lambda_{\text{max}}<0)$] varying with
$\sigma^2$. Increasing $m$ results in the transition to instability
occurring at higher $\sigma^2$, implying that network stability
increases with modularity.}
\label{fig:star_modular}
\end{figure}
The above results were obtained for a specific value of $L$ ($=N-1$).
We now relax the constraint on link cost and allow a larger
number of links than that strictly necessary to keep the network connected.
The larger $L$ is manifested as random links between
nonhub nodes, resulting in higher clustering within the network. Even for such
{\it clustered star} networks, $\lambda_{\text{max}}$  increases with
size as $\sqrt{N}$, and therefore, their instability is reduced by imposing
a modular structure (Fig.~\ref{fig:star_modular}).  The effect of
increasing the number of modules, $m$, on the dynamical stability of a
network can be observed from the stability-instability transition that
occurs on increasing the network parameter $\sigma$ keeping $N, C$ fixed.
The critical value at which the transition to instability
occurs, $\sigma_c$, increases with $m$ (Fig.~\ref{fig:star_modular}, inset)
while $\ell$ does not change significantly. This signifies that even for
large $L$, networks satisfy the structural and
functional constraints by adopting a modular configuration.

\begin{figure}
\begin{center}
\includegraphics[width=0.9\linewidth]{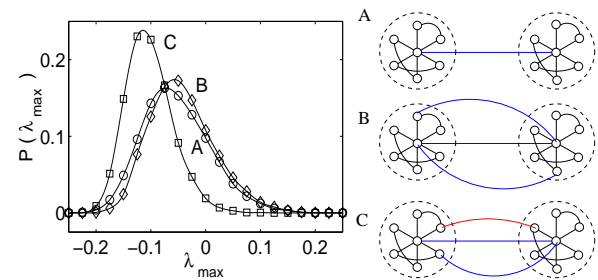}
\end{center}
\caption{Probability distribution of $\lambda_{\text{max}}$ for clustered
star networks ($N = 256, L = 15 N$) having four modules with
different types of intermodular connectivities (A), (B) and (C), which
are represented schematically here.
Link weights $W_{ij}$ have a $Normal~(0,\sigma^2)$
distribution with $\sigma^2=0.018$.}
\label{fig_different}
\end{figure}
As $L$ is increased, we observe that the additional links in the
optimized network occur between modules, in preference to, between nodes in
the same module. To see why the network prefers the former configuration,
we consider three different types of intermodular connections: (A)~only the
hub nodes of different modules are connected, (B)~nonhub nodes of one
module can connect to the hub of another module, and (C)~nonhub nodes of
{\em different} modules are connected.  Arrangement (B) where intermodular
connections that link to hubs of other modules actually increase the
maximum degree in the modules, making this arrangement more unstable than
(A).  On the other hand, (C) connections between nonhub nodes of {\em
different} modules not only decrease the instability
(Fig.~\ref{fig_different}), but also reduce $\ell$.  As a result, the
optimal network will always prefer this arrangement (C) of large number of
random intermodular connections over other topologies for large
$L$.

\begin{figure}
\begin{center} 
  \includegraphics[width=0.9\linewidth]{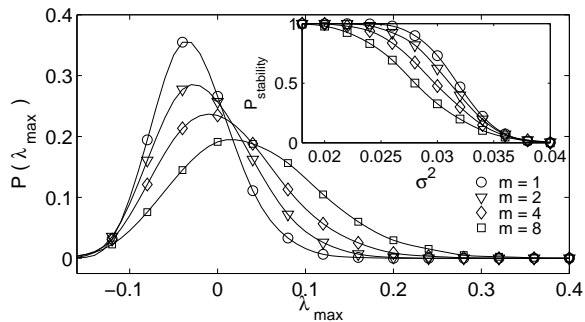}
\end{center}
\caption{Probability distribution of $\lambda_{\text{max}}$ for random
networks ($N=256, L = 15 N$) as a function of the number of
modules, $m$, which are connected to each other by single links.  Link
weights $W_{ij}$ follow $Normal~(0,\sigma^2)$ distribution with
$\sigma^2=0.03$. The inset shows the probability of stability
[$P(\lambda_{\text{max}} < 0)$] varying with $\sigma^2$. Increasing $m$
results in transition to instability at lower $\sigma^2$,
indicating that increasing modularity decreases stability for random
networks.}
  \label{random_modularity}
\end{figure}

Our observation that {\em both} structural and dynamical constraints are
necessary for modularity to emerge runs counter to the general belief that
modularity necessarily follows from the requirement of robustness {\em
alone}, as modules are thought to limit the effects of local perturbations
in a network.  To further demonstrate that the three constraints are the
minimal required for a network to adopt a modular configuration, we remove
the hub from a clustered star while ensuring that the network is still
connected. This corresponds to the absence of the link cost constraint
altogether and the optimal graph is now essentially a random network.  To
see why modularity is no longer observed in this case, we consider the
stability of an ER random network on which a modular structure has been
imposed.  A network of $N$ nodes is divided into $m$ modules, connected to
each other with a few intermodular links.  We then consider the
stability-instability transition of networks for increasing $m$, with the
average degree, $\langle k \rangle$ kept fixed.  Although from the
May-Wigner theorem, it may be naively expected that $\sigma_c \simeq
1/\sqrt{\langle k \rangle}$ is constant w.r.t. $m$, we actually observe
that increasing $m$ decreases stability (Fig.~\ref{random_modularity}).  
This is because when a network of size $N$ is split into $m$ modules, the
stability of the entire network is decided by that of the most unstable
module, ignoring the small additional effect of intermodular connections.
Thus, the stability of the entire network is decided by randomly drawing
$m$ values from the distribution of $\lambda_{\text{max}}$ for the modules.
Therefore, for modular networks it is more likely that a positive
$\lambda_{\text{max}}$ will occur,
than for the case of a homogeneous random network of size
$N$~\cite{Hastings92}.  The {\em decrease} of stability with modularity for
random networks shows that, in general, it is not necessary that modularity
is always stabilizing and results in a robust network, as has sometimes
been claimed~\cite{Variano04}.

In this paper we have shown that modules of interconnected nodes can arise
as a result of optimizing between multiple structural and functional
constraints. In particular, we show that by minimizing link cost as well as
path length, while at the same time increasing robustness to dynamical
perturbations, a network will evolve to a configuration having
multiple modules characterized by hubs, that are connected to each other.
At the limit of extremely small $L$ this results in networks with
bimodal degree distribution, that has been previously shown to be robust
with respect to both targeted and random removal of
nodes~\cite{Tanizawa05}. Therefore, not only are such modular networks
dynamically less unstable, but they are also robust with respect to structural
perturbations. In general, on allowing larger $L$, the optimized
networks show heterogeneous degree distribution that has been observed in a
large class of networks occurring in the natural and social world,
including those termed as scale free networks~\cite{Albert02}. Our results
provide a glimpse into how the topological structure of complex networks
can be related to functional and evolutionary considerations.

We thank R.~Anishetty and S.~Jain for discussions.

\end{document}